\documentclass[a4paper,12pt]{article}
\usepackage{amsmath,amsthm,amssymb,bbm}

\usepackage[]{inputenc}
\usepackage[T1]{fontenc}
\usepackage[english]{babel}
\usepackage[]{csquotes}
\usepackage[]{graphicx}
\usepackage{subfigure}
\usepackage[]{cite}
\usepackage{amssymb}
\usepackage{bbm}
\usepackage{braket}

\title{\bf Pulsed homodyne Gaussian quantum tomography with low detection efficiency}

\author{M. Esposito$^1$, F. Benatti$^{1,2}$, R. Floreanini$^2$, S. Olivares$^3$, F. Randi$^1$, K. Titimbo$^1$,\\
 M. Pividori$^1$, F. Novelli$^4$, F. Cilento$^4$, F. Parmigiani$^{1,4}$ and D. Fausti$^{1,4}$\\
\\
\small ${}^1$Dipartimento di Fisica, Universit\`a di Trieste, 
34127 Trieste, Italy\\
\small ${}^2$Istituto Nazionale di Fisica Nucleare, Sezione di Trieste,
34014 Trieste, Italy\\
\small ${}^3$Dipartimento di Fisica, Universit\`{a} degli Studi di Milano, 20133 Milano, Italy\\
\small ${}^4$Sincrotrone Trieste S.C.p.A., 34127 Basovizza, Italy
}

\date{\null}

\begin{document}

\maketitle

\begin{abstract}
\noindent
Pulsed homodyne quantum tomography usually requires a high detection efficiency
limiting its applicability in quantum optics. Here, it is shown that the presence of low
detection efficiency ($<50\%$) does not prevent the tomographic reconstruction of quantum states of light, specifically, of Gaussian type. This result is obtained by applying
the so-called ``minimax'' adaptive reconstruction of the Wigner function to pulsed homodyne detection. In particular, we prove, by both numerical and real experiments,
that an effective discrimination of different Gaussian quantum states can be achieved.
Our finding paves the way to a more extensive use of quantum tomographic methods,
even in physical situations in which high detection efficiency is unattainable.
\end{abstract}

\section{Introduction}

Standard homodyne detection is an experimental method that is used for the reconstruction of quantum states of monochromatic light. In this framework, the quantum state is characterized by feeding to appropriate tomografic techniques the repeated measurement of a discrete set of field quadratures \cite{Vogel,Smith,Welsch}. Quantum state reconstruction methods turn out to be of paramount importance for quantum information, for they can reveal the presence of quantum coherence and entanglement, not possible in a classical setting \cite{Nielsen}.

For these experiments, a very high detection efficiency is tipically required, along with ad-hoc designed apparatus \cite{Zavatta}. However, new methods capable of discriminating between different quantum state of light, which demand lower detection capabilities typical of commercially available components, would facilitate the experimental realization of such apparatuses. Moreover, it would make possible to apply quantum homodyne detection to study different physical systems where high noise conditions are unavoidable, such as out of equilibrium light matter dynamics \cite{PP}. For this purpose, novel quantum state reconstruction methods should be considered. Here it is proved that even in high noise conditions (equivalent detection efficiency $<50\%$) a proper reconstruction of the state of light can be efficiently achieved by ``minimax'' adaptive estimation of the Wigner function \cite{Butucea,Aubry}. This statistical technique is able to overcome the difficulties that arise when the more standard pattern function based quantum tomography is adopted \cite{DAriano1,Kiss1,Herzog,DAriano2,Kiss2,Richter,DAriano3,Lvovsky}.

In this manuscript, we report on the commissioning and characterization of a time-domain homodyne detection apparatus working with coherent ultra-short light pulses, built using commercial detectors and operating in a regime of large electronic noise. By treating the shot-to-electronic-noise ratio (about 2 dB) as a detector inefficiency \cite{Leonhardt,Appel},
we obtain an overall detection efficiency of about $30\%$.
By extending single mode tomographic techniques to the pulsed regime, we show that, even in these low efficiency conditions, it is still possible to discriminate coherent and squeezed photon states with high accuracy.

The effectiveness of these tomographic methods has been tested by numerical experiments and for the tomographic reconstruction of coherent states with different mean photon number in real experiments.

\section{Homodyne Detection}

In standard single-mode homodyne detection, the photon state under investigation, the \emph{signal}, is mixed with a strong coherent reference state, the local oscillator (\emph{LO}), by a $50/50$ beam splitter. The outputs are collected by two photodiodes and the difference photocurrent (homodyne photocurrent) is measured. It can be proven that, when the \emph{LO} is significantly more intense than the \emph{signal}, the homodyne photocurrent is proportional to the \emph{signal} quadrature \cite{Ferraro}.
Denoting by $\hat{a}$ and $\hat{a}^{\dagger}$ the single mode annihilation and creation operators associated with the \emph{signal}, the quadrature operator is defined as,
\begin{eqnarray}
    \hat{x}_{\Phi}=\frac{ \hat{a} e^{-i \Phi} + \hat{a}^{\dagger}e^{i \Phi}}{\sqrt{2}}\ ,
		 \label{quad}
\end{eqnarray}
where $\Phi$ is the relative phase between the \emph{signal} and the \emph{LO}. The continuum set of quadratures with $\Phi \in [0, 2\pi]$ provides a complete characterization of the \emph{signal} state. 

The standard techniques developed for the reconstruction of single-mode photon states need to be generalized when pulsed light is used \cite{Hansen,Zavatta2,Kumar,Haderka,Cooper,Chi,Okubo}. In this case, the optical pulses are equally prepared multi-mode coherent quantum states. For each single pulse a quadrature measurement is performed. In such regime, the single-mode description can still be used if the single-mode field operators are replaced by multi-mode annihilation and creation operators. For this reason we make use in the following of the single-mode notation. A thorough theoretical treatment for pulsed homodyne detection is presented in the Appendix.

\subsection{Experimental setup}

Fig.\ref{Fig1} shows the experimental set-up.
\begin{figure}[htbp]
 \centering
   \includegraphics[width=0.6\textwidth]{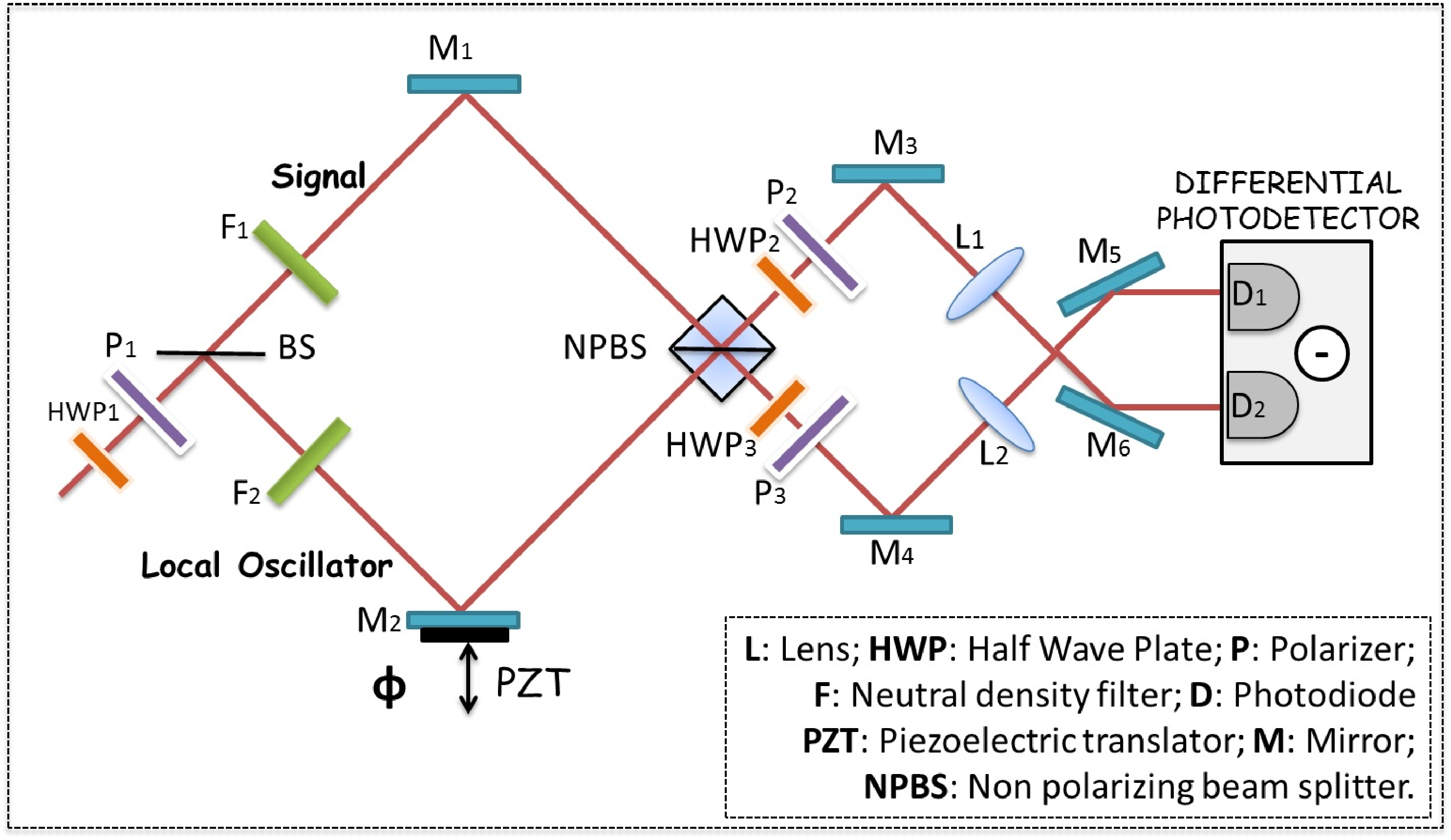}
  \caption{\small Scheme of the opto-mechanical setup.}
   \label{Fig1}
\end{figure}
The laser source is a mode-locked Ti:Sapphire oscillator with $80$ MHz repetition rate. A beam splitter divides the incoming beam in two parts which then interfere in a second beam splitter (NPBS in Fig.\ref{Fig1}). The outputs are detected and subtracted by a commercial differential photodetector. The latter is made up of two $Si$/PIN photodiodes with nominal quantum efficiency $\eta_{pd}=0.85$ at $800$ nm wavelength and linear response up to $0.6$ mW \emph{LO} power. The detector subtraction efficiency is quantified by the common mode rejection ratio (CMRR), defined as the ratio between the detector output power when both photodiodes are illuminated and the power when one of the two is screened \cite{Chi}. For the present experiment CMRR$>36$ dB.

The homodyne photocurrent is recorded by a digital oscilloscope with a bandwidth of $500$ MHz and a sampling rate of $5$ GSamples/s. The digitized output is numerically integrated over time intervals corresponding to the duration of the pulse. Each integral is associated with a single quadrature measurement.

\begin{figure}[hbt]
\centering
 \includegraphics[width=0.6\textwidth]{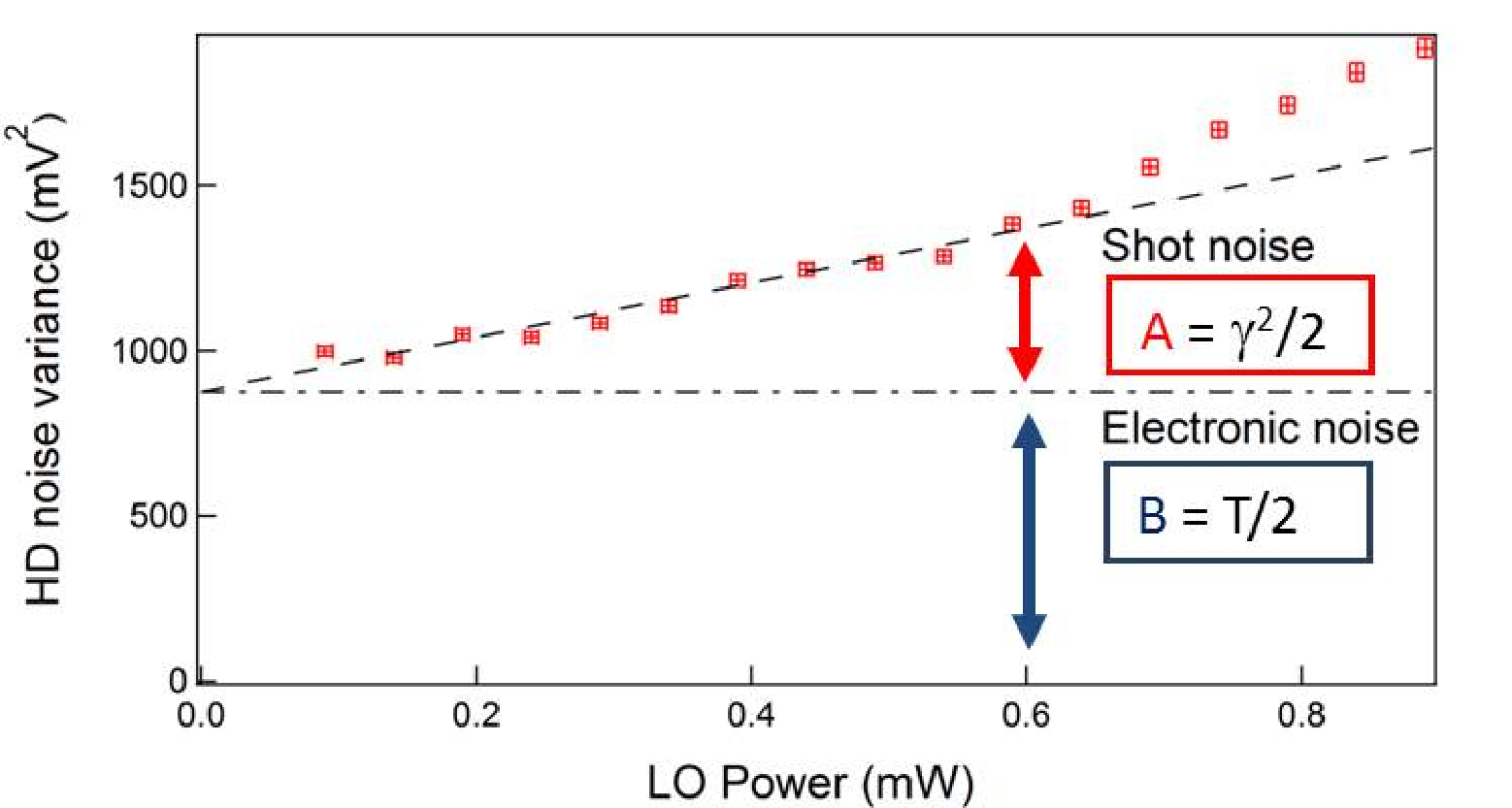}
  \caption{\small Detector noise variance versus \emph{LO} power in absence of signal (vacuum state); shot noise contribution  (dashed curve); electronic noise background (dashed-dotted curve).}
  \label{Fig2}
\end{figure}

In the \emph{shot noise regime} with the \emph{signal} beam blocked, \textit{i.e.} with the \emph{signal} in the vacuum state, the homodyne detector noise variance is expected to change linearly with the \emph{LO} power on the top of a constant offset representing the electronic noise \cite{Bachor}. Fig.\ref{Fig2} shows the detector noise variance of $8 \times 10^3$ pulses, for different values of the \emph{LO} power. The noise variance grows linearly up to $0.6$ mW \emph{LO} power, instead at higher powers the photodiodes non-linear effects are significant.  To have the maximum shot-to-electronic noise ratio ($\approx 2$ dB) achievable in the linear regime, the experiments have been performed at $0.6$ mW \emph{LO} power.
\\
Conversely, when the \emph{signal} beam is not blocked, it can be attenuated with respect to the \emph{LO} by neutral density filter ($F_1$ in Fig.\ref{Fig1}). The phase difference $\Phi$ between the two arms of the interferometer can be modulated in the $[0, 2\pi]$ range using a piezoelectric translator in the \emph{LO} arm ($PZT$ in Fig.\ref{Fig1}). The homodyne photocurrent is acquired by the digital oscilloscope and each difference pulse is integrated.

\subsection{Homodyne traces}

In the absence of electronic noise, the voltage $V$ corresponding to the homodyne photocurrent,
namely the experimentally accessible quantity, is proportional to the quadrature operator $x_{\Phi}$,
$V=\gamma\,x_{\Phi}$, with an appropriate constant $\gamma$.
The electronic noise can be generically described by a classical stochastic process $\delta$, that
can be assumed to be Gaussian distributed, with zero average and variance $T/2$. The value of $\delta$ must be experimentally measured. Under these conditions the electronic noise is independent from the quadrature and it contributes to the homodyne voltage as, 
\begin{eqnarray}
  V = \gamma \, x_{\Phi} + \delta\ .
  \label{prop}
\end{eqnarray}
A homodyne trace is obtained by collecting a set of homodyne voltage values $V_i$, corresponding to different phase values $\Phi_i$, associated with a large number $M$ of piezo positions.

In the case of the vacuum state (absence of \emph{signal}), these considerations are outlined in Fig.\ref{Fig2}. The total vacuum homodyne variance for the chosen \emph{LO} power is indeed composed of two independent contributions. An intrinsic contribution ($A=\gamma^2/2$) is actually the shot noise, while an extrinsic contribution ($B=T/2$) is due to the electronic noise.
Following Ref.\cite{Appel}, the electronic noise can be treated as an optical loss channel with an equivalent transmission efficiency given by
\begin{eqnarray}
  \eta_{eq} = \frac{A}{A+B} = \frac{\gamma^2}{{\gamma'}^2}\ ,
    \label{etaeq}
\end{eqnarray}
where $\gamma' = \sqrt{\gamma^2 + T}$. Here we estimate $\eta_{eq} = 0.37$. Thus the total apparatus efficiency is $\eta = \eta_{eq}\, \eta_{pd}=0.31$.

Considering that the variances of independent stochastic variables are additive, $\gamma'$ can be determined by using the vacuum state as reference and by assuming the variance of the quadrature operator in the vacuum to be  $1/2$,
\begin{eqnarray}
\gamma' = \sqrt{2  \braket{V^2}_0}\ ,
  \label{gamma}
\end{eqnarray}
where $\braket{V^2}_0$ is the experimental voltage variance for the vacuum state.

\begin{figure}[hbt]
\centering
\subfigure[\label{Fig3aNew}]{\includegraphics[height=4.5cm]{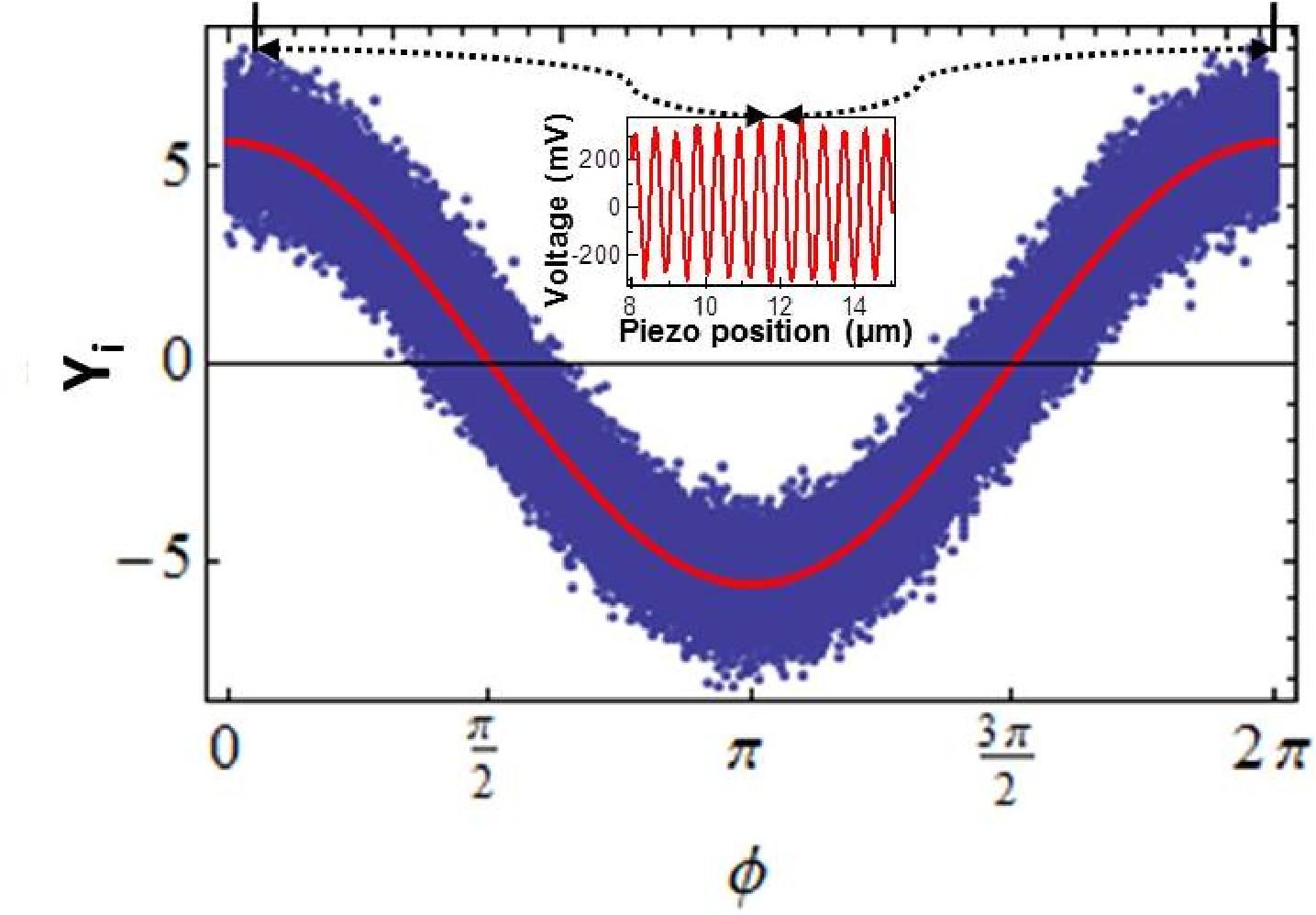}}
\quad
\subfigure[\label{Fig3b}]{\includegraphics[height=4.5cm]{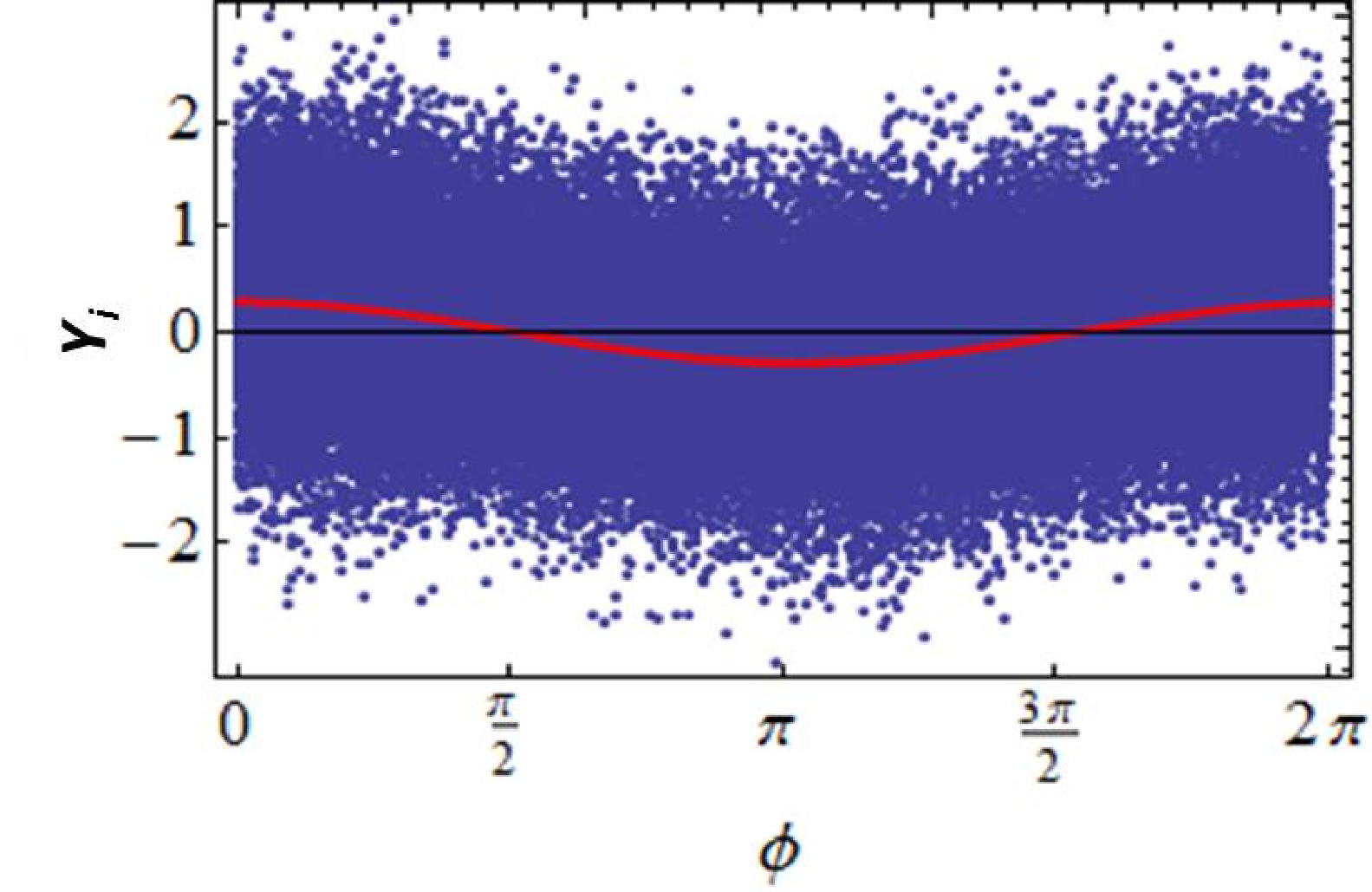}}
\caption{\small Calibrated homodyne traces of two optical coherent states. Each trace is acquired using a different optical density $OD$ of the filter $F_1$.  (a) $OD = 4.5$; (b) $OD = 7$. In the inset of (a) we show the interference figure obtained by measuring the mean value of four integrated pulses versus the piezo position. The homodyne traces are measured in the piezo range corresponding to the central optical cycle of the interference figure.}
 \label{Fig3}
\end{figure}

To consistently apply the quantum state reconstruction to the collected experimental homodyne data, it is convenient to rescale the raw data $V_i$ by the constant $\gamma'$,
\begin{eqnarray}
Y_i = V_i/\gamma' \ ,
 \label{calibrated}
\end{eqnarray}
so that the new, calibrated quantities $Y_i$ have variance $1/2$. 
Fig.\ref{Fig3} shows the case of two optical coherent states with different mean photon numbers, each homodyne trace consisting of $M=8 \times 10^4$ experimental data.
The data were collected using two different optical density for the signal attenuator (see $F_1$ in Fig.\ref{Fig1}).
Clearly even with a very low mean photon number signal (see Fig.\ref{Fig3b}), the phase modulation of the homodyne trace
is detectable.

\section{Quantum state reconstruction}

In the following, we prove that a minimax estimation of the Wigner function \cite{Butucea} can discriminate between quantum coherent and squeezed states even under the low efficiency condition of our experiment.

\subsection{Minimax estimation of the Wigner function}

From the photon operators $\hat{a},\hat{a}^\dag$ one constructs position-like, $\hat{q}=(\hat{a}+\hat {a}^\dag)/\sqrt{2}$, and momentum-like operators,  $\hat{p}=(\hat{a}-\hat {a}^\dag)/(i\sqrt{2})$. Given a density matrix $\hat{\rho}$ representing a generic photon state, the associated Wigner function $W_\rho(q,p)$ in the variables $(q,p)$ is defined by,
\begin{equation}
W_\rho(q,p)=\frac{1}{(2\pi)^2}\int_{\mathbb{R}^2}{\rm d}u\, {\rm d}v\, {\rm e}^{i(uq+vp)}\, {\rm Tr}\Big[\hat{\rho}\,{\rm e}^{-i(u\hat{q}+v\hat{p})}\Big]\ .
\label{wigner}
\end{equation}
The mean value of any observable $\hat{O}$, expressible as a function of $\hat{a}$ and $\hat{a}^\dagger$, can be written as,
\begin{eqnarray}
\langle \hat{O}\rangle_\rho={\rm Tr}(\hat{\rho}\,\hat{O})=\int_{\mathbb{R}^2}dq\,dp\,{W}_\rho(q,p)\, {O}(q,p)\ ,
  \label{Wigner}
\end{eqnarray}
where the function $O(q,p)$ is connected to the operator $\hat{O}$ as the Wigner function is to $\hat{\rho}$
as in (\ref{wigner}), provided one changes the signs of the two exponents.
The Wigner function $W_\rho(q,p)$ is therefore sufficient to evaluate the average value of any observable $\hat{O}$ with respect to any state $\hat{\rho}$.

The minimax reconstruction algorithm precisely gives an estimator for the Wigner function, allowing its reconstruction from the collected homodyne data $Y_i$, proviso a proper calibration is used Eq. \ref{calibrated}. The estimator is given by,
\begin{eqnarray}
{W}^{\eta}_{h, M} (q, p) = \frac{1}{M} \sum_{i=1}^{M} K^{h,\eta}_{Y_i,\Phi_i}(q,p)\ ,
\label{wigEstimator}
\end{eqnarray}
and it depends on the overall detector efficiency $\eta$, the number of collected data $M$ and an adaptive parameter $h$ \cite{Butucea}. The kernel function
$K^{h,\eta}_{Y_i,\Phi_i}(q,p)$ is explicitly given by
\begin{eqnarray}
K^{h,\eta}_{Y_i,\Phi_i}(q,p) = \int_{-\frac{1}{h}}^{\frac{1}{h}} dt \frac{|t|}{4 \pi} e^{- \imath t (q \cos{\Phi_i} + p \sin{\Phi_i} - \frac{Y_i}{\sqrt{\eta}})+ t^2 \frac{1-\eta}{4 \eta}}\ .
\label{kernel}
\end{eqnarray}
The truncation of the integration due to the parameter $h$ is necessary because of the diverging inverse Gaussian present in the integral. In Ref.\cite{Butucea} it is  shown that the optimal adaptive estimator, {\it i.e.} the one that minimizes uncertainties, is obtained when $h=h_{adap}$, with
\begin{eqnarray}
h_{adap} = \left(\frac{2\eta \log{M}}{1-\eta}- \sqrt{\frac{2 \eta \log{M}}{1-\eta}}\right)^{-1/2}\ .
\label{hadap}
\end{eqnarray}
Notice that the optimally adapted cutoff depends both on the detector efficiency $\eta$ and the number $M$ of collected data. When the efficiency becomes small, in order to have a good estimator of the Wigner function, a larger number $M$ of data is needed. However, we stress that no inferior bound on $\eta$ exists that prevents the convergence of the algorithm.

Once the Wigner function is reconstructed, the expectation value of any observable $\hat{O}$ of the system can be estimated as follows,
\begin{equation}
\label{expectationValue0}
E^{h}_{\hat{O}}= \int_{\mathbb{R}^2} dq \, dp \  {O}(q, p) \, {W}^{\eta}_{h, M} (q, p)=\frac{1}{M} \sum_{i=1}^{M} R^{h,\eta}_{\hat{O}}(Y_i, \Phi_i)\ ,
\end{equation}
where the corresponding kernel function is given by
\begin{equation}
\label{opstimator2}
R^{h,\eta}_{\hat{O}}(Y_i, \Phi_i)=\int_{\mathbb{R}^2} dq \, dp \ {O}(q,p) \, K^{h,\eta}_{Y_i,\Phi_i}(q,p) \ .
\end{equation}

\subsection{Numerical experiments}

For testing the effectiveness of this approach under our experimental conditions, we first analyze sets of numerically generated data that simulate the quadratures of a known pure quantum state $\vert\psi\rangle$. By knowing in advance the quantum state, it is possible to verify whether the minimax technique allows a proper reconstruction.

The generation of the fictitious data starts by using the quadrature probability distribution with $\eta=1$  associated with a quantum state, and afterwards in adding to each numerically generated state quadrature a Gaussian noise which exactly simulates the electronic noise associated with the efficiency, \textit{i.e.} $\eta=0.31$, of our set-up. The analysis of the simulated data proceeds, as for experimental data, through a calibration using the vacuum noise as reference.
In particular, we perform two numerical experiments. The first with a coherent state $\vert\psi\rangle=D(\alpha) \ket{0}$ and another one with a displaced-squeezed state $\vert\psi\rangle=D(\alpha) S(\xi) \ket {0}$, where 
\begin{equation}
D(\alpha)=e^{\alpha \hat{a}^{\dagger}- \alpha^{\ast} \hat{a}}\ ,\qquad \ S(\xi)=e^{1/2(\xi \hat{a}^{\dagger 2}- \xi^{\ast} \hat{a}^2)}\ ,
\end{equation}
are the displacement and squeezing operators, respectively.
Each data set consists of\break \hbox{$M=8 \times 10^4$} quadrature measurements with the relative phase $\Phi$ ranging in the interval 
$[0, 2\pi]$.

\begin{figure}[hbt]
\centering
\subfigure[\label{Fig4a}]{\includegraphics[width=0.45\textwidth]{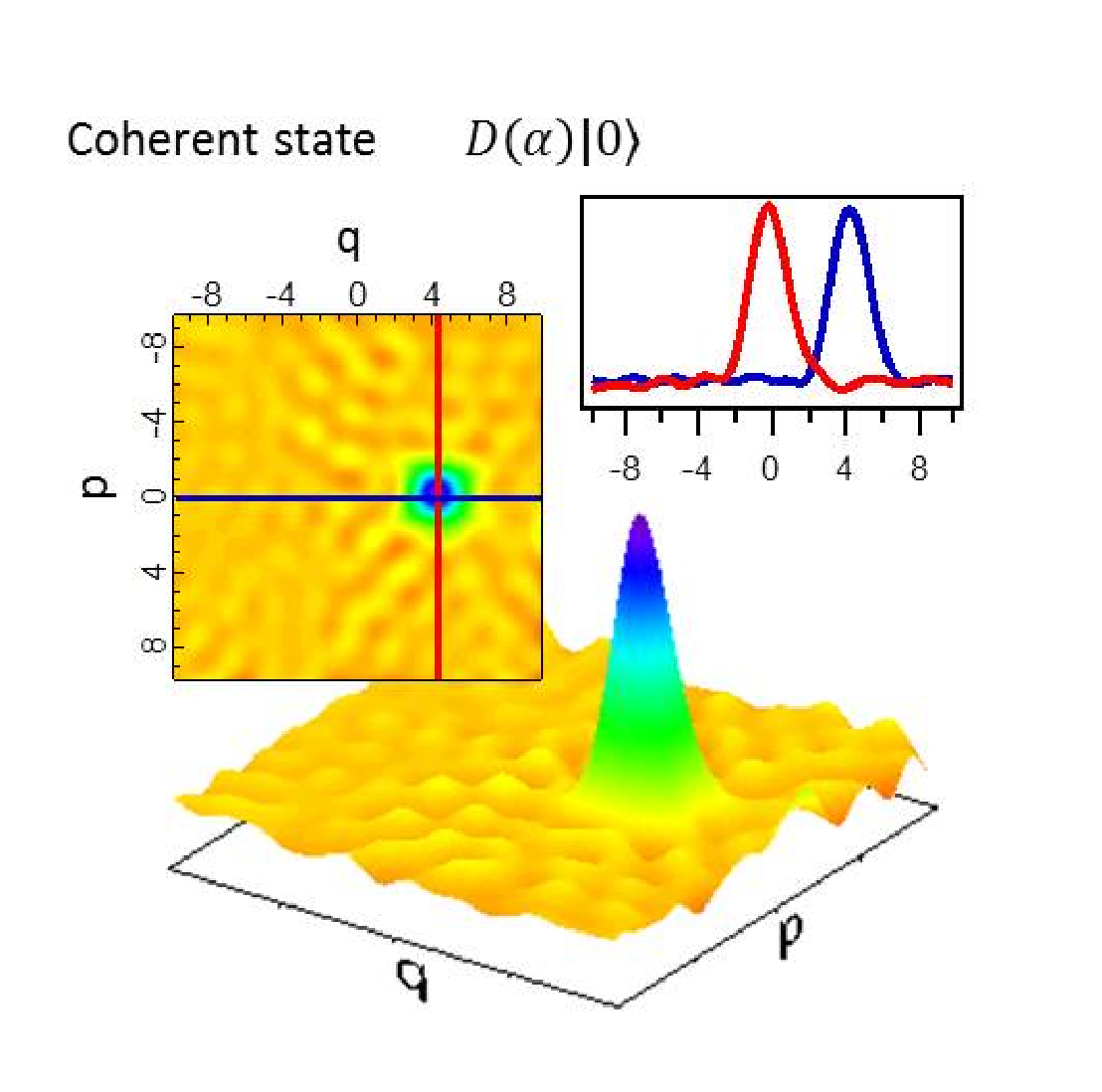}}
\quad
\subfigure[\label{Fig4b}]{\includegraphics[width=0.45\textwidth]{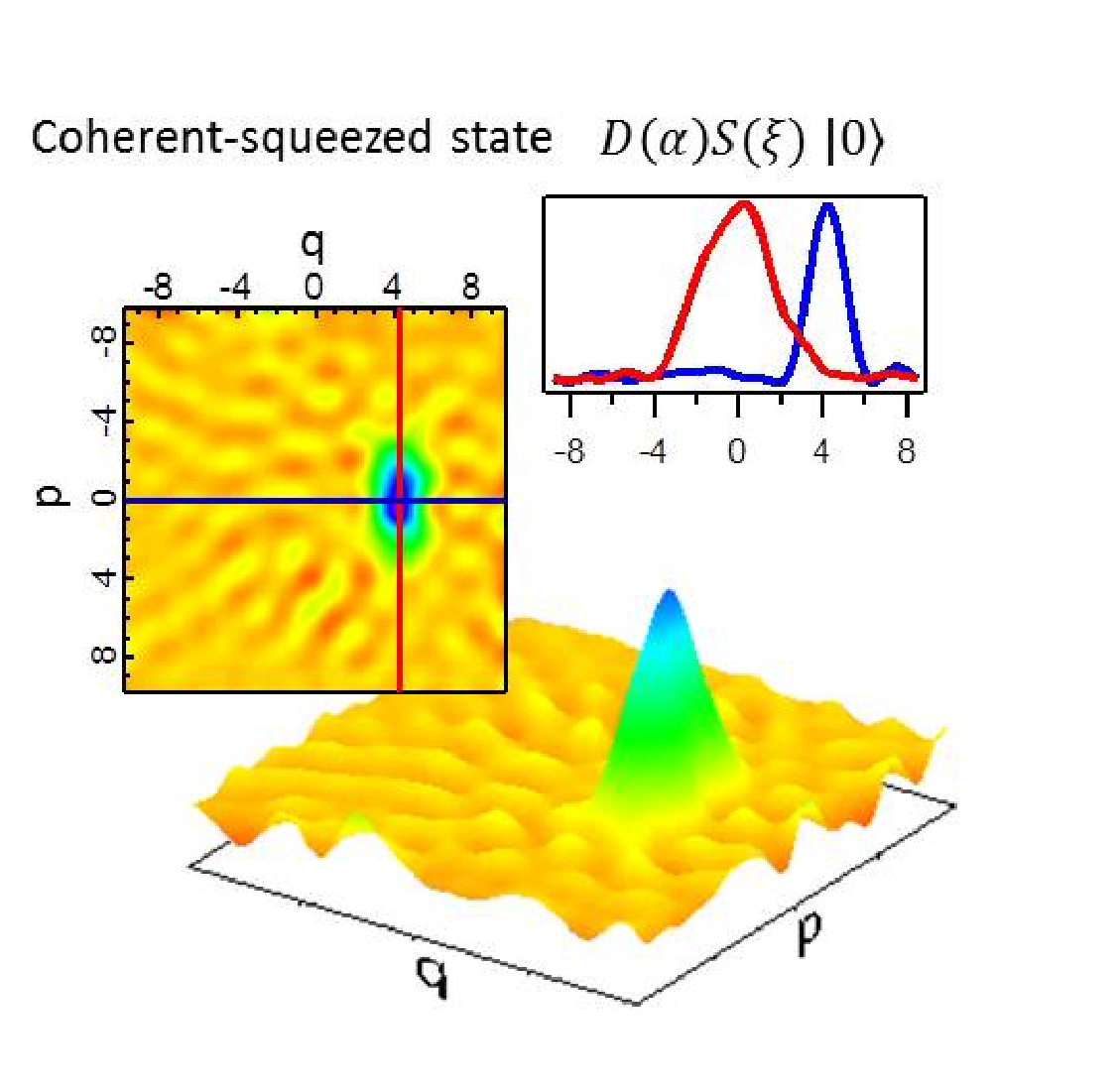}}
\caption{\small Reconstructed Wigner function using the adaptive minimax estimation from $M=8 \times 10^4$ quadratures of (a) a coherent state ($\alpha=3$) and (b) a displaced-squeezed state ($\alpha=3$, $\xi=0.8$). The used detector efficiency is in both cases $\eta=0.31$. The inserts show representative two-dimensional cuts of the reconstructed Wigner function.}
 \label{Fig4}
\end{figure}

Following the algorithm sketched before, we reconstruct the Wigner function of the two quantum states. They are shown in Fig.\ref{Fig4} (a) and (b) for the coherent and the squeezed state, respectively.
To remove the artifacts resulting by the numerical integration, the reported Wigner functions have been filtered by an image processing algorithm (low pass Gaussian convolution filtering). The blue and red curves are bidimensional cuts of the Wigner functions in correspondence of the expectation value of the position and momentum operator, respectively. The fidelity of the Wigner function reconstruction has been calculated as follows,
\begin{equation}
\label{fidelity}
f=\int_{\mathbb{R}^2} dq \, dp \ 2 \pi \ W_e(q,p) \, W_r(q,p) \ ,
\end{equation}
where $W_e(q,p)$ is the exact Wigner function and $W_r(q,p)$ is the reconstructed one. We obtained a fidelity of $0.97$ and $0.92$ for the coherent and the squeezed state respectively.
The results prove that the squeezed and coherent states can be discriminated under high noise conditions. Indeed, the features of the reconstructed Wigner functions clearly reflect the different nature of the two quantum states.

From the reconstructed Wigner function, it is possible to compute the expectation values of relevant observables with respect to the coherent and the displaced-squeezed state.
In particular, the number operator $\hat{n}=\hat{a}^{\dagger} \hat{a}$ and the position $\hat q$ and momentum $\hat{p}$ operators. For $\hat q$ and $\hat{p}$ it also possible to derive the variances, $\sigma^2 [\hat{q}]$ and $\sigma^2 [\hat{p}]$, and to estimate the squeezing parameter,
\begin{equation}
\xi=\frac{1}{4}\ln{(\sigma_{\hat{p}}^2/\sigma_{\hat{q}}^2)}\ .
\end{equation}
The results are summarized in Table\ref{tabSimulatedCh}. 
For the errors evaluation we have used the standard expression for the mean average 
error relative to a data set $(Y_i,\Phi_i)$ (recall the definitions (\ref{expectationValue0}) and (\ref{opstimator2})),
\begin{eqnarray}
\epsilon_{\hat O} = \sqrt{\frac{\sum_{i=1}^{M} \big[R^{h,\eta}_{\hat{O}}(Y_{i}; \Phi_i )- E_{\hat{O}}\big]^2}{M(M-1)}}\ .
    \label{err}
\end{eqnarray}
The averages computed using the Wigner function estimated from the data set fully agree with those analytically calculated from the known quantum states.
In particular, even the estimated value of the squeezing parameter agrees with the expected one.
This reveals that the minimax reconstruction method is effective in discriminating different quantum states even under low efficiency conditions.
\renewcommand{\arraystretch}{1.1}
\begin{table}[hbt]
 \caption{\small Estimate of different expectation values for $M=8 \times 10^4$ simulated quadratures associated to the coherent state ($\alpha=3$) and the displaced-squeezed state ($\alpha=3$, $\xi=0.8$).}
\begin{center}
\begin{tabular}{| c || c | c || c | c |}
   $\hat{O}$ & $\braket{\hat{O}}_{D(\alpha) \ket {0}}$ & $E^h_{\hat{O}}$ & 
   $\braket{\hat{O}}_{D(\alpha) S(\xi) \ket {0}}$ & $E^h_{\hat{O}}$\phantom{$\bigg|$}\\ \hline
   $\hat{n}$ & $9$ & $9 \pm 1$ & $9.89$ & $9.9 \pm 0.8$\\
   $\hat{q}$ & $4.24$ & $4.2 \pm 0.5$ & $4.24$ & $4.3 \pm 0.3$\\
   $\hat{p}$ & $0$ & $0.00 \pm 0.06$ & $0$ & $0.0 \pm 0.2$\\
   $\sigma_{\hat{q}}^2$ & $0.5$ & $0.4 \pm 0.2$ & $0.1$ & $0.1 \pm 0.2$\\
   $\sigma_{\hat{p}}^2$ & $0.5$ & $0.5 \pm 0.2$ & $2.48$ & $2.3 \pm 0.4$\\
   $\xi$ & $0$ & $0.0\pm 0.1$ & $0.8$ & $0.78 \pm 0.02$\\
\end{tabular}
\end{center}
\label{tabSimulatedCh}
\end{table}

\subsection{Real Experiments}

\begin{figure}[hbt]
\centering
\subfigure[\label{Fig5a}]{\includegraphics[width=0.45\textwidth]{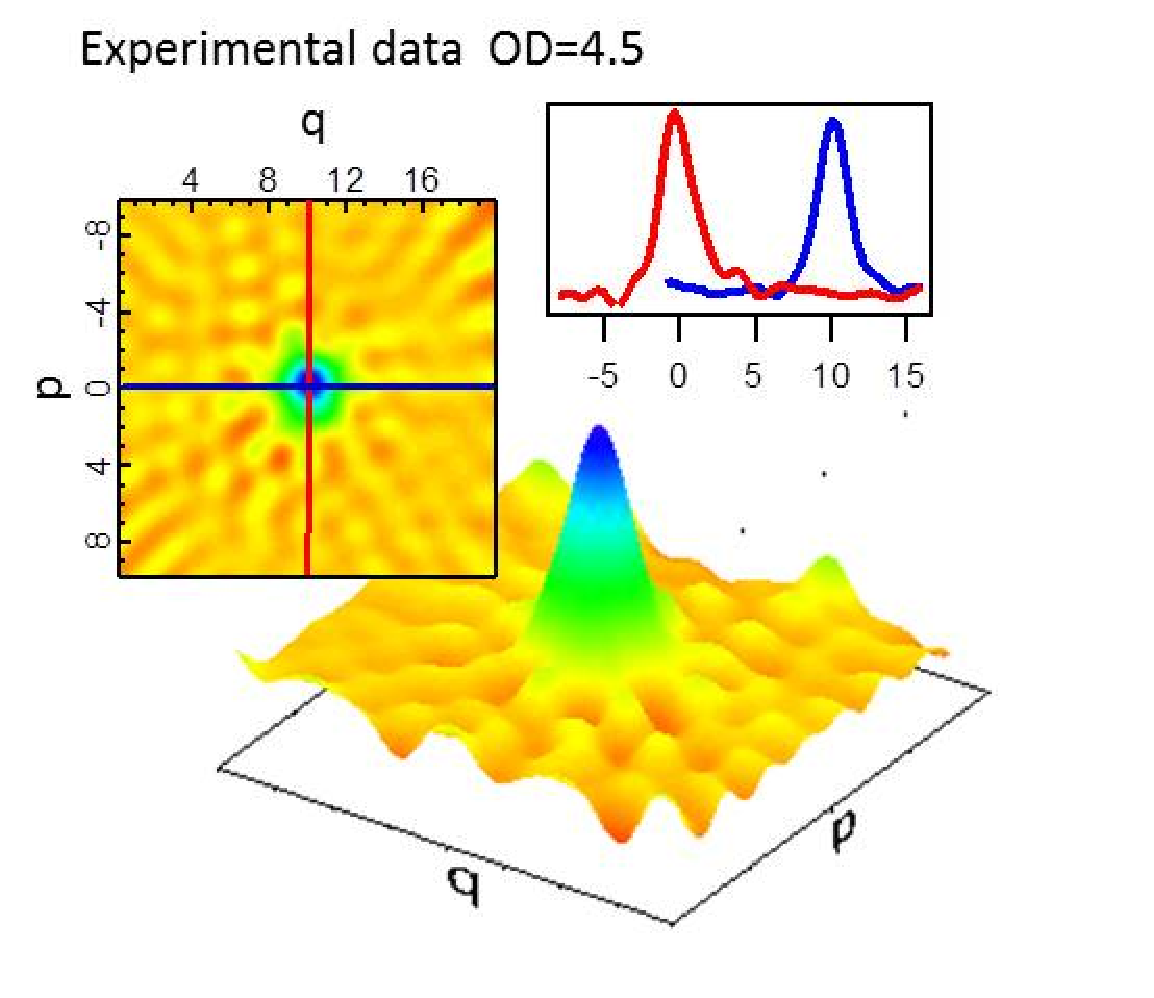}}
\quad
\subfigure[\label{Fig5b}]{\includegraphics[width=0.45\textwidth]{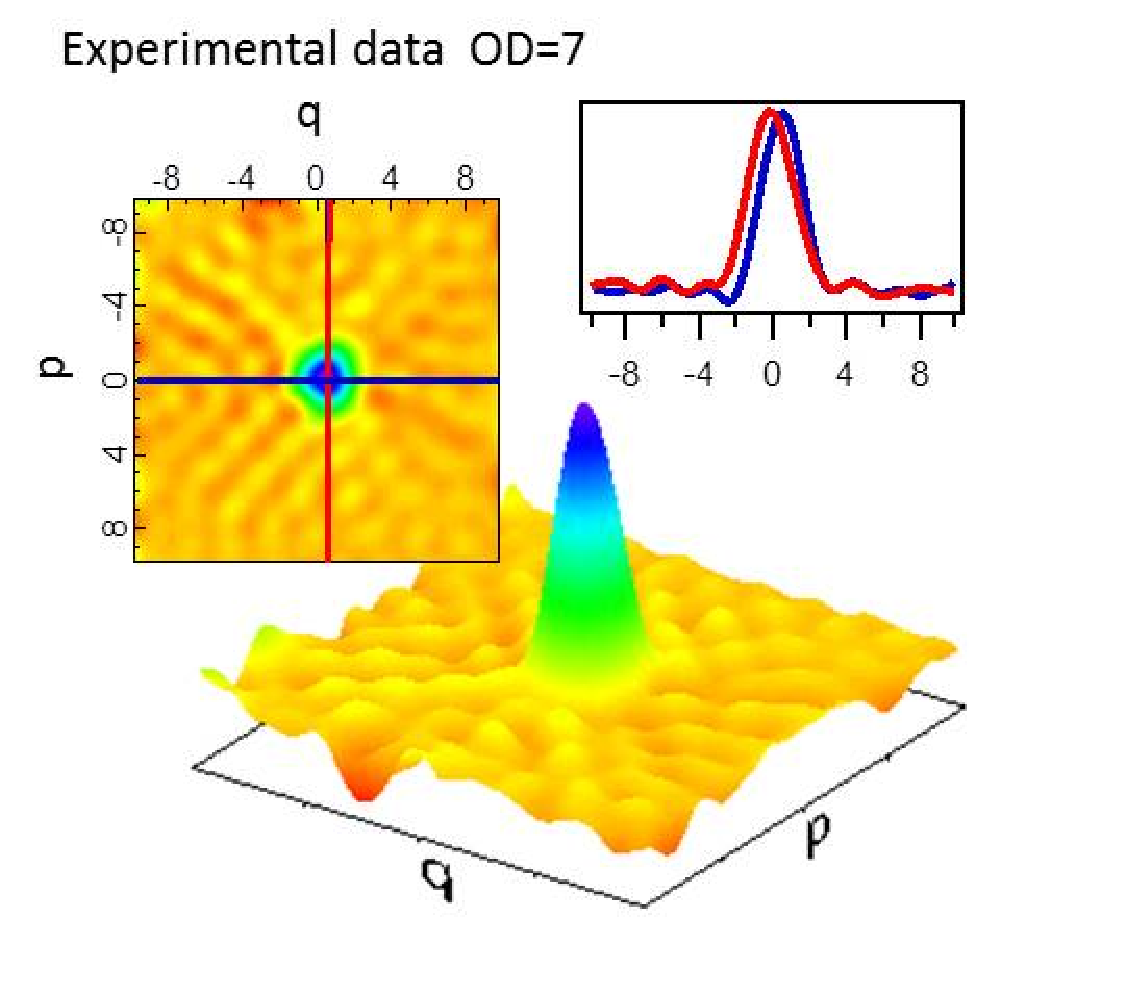}}
\caption{\small Reconstructed Wigner function using the adaptive minimax estimation from $M=8 \times 10^4$ quadratures of attenuated laser states with (a) $OD=4.5$ and (b) $OD=7$. The inserts show representative two-dimensional cuts of the reconstructed Wigner function.}
 \label{Fig5}
\end{figure}

The real data in Fig.\ref{Fig3} can now be treated in the same way.
In this case, the collected homodyne traces are two coherent states with very different mean photon number.\\
The first step in the reconstruction procedure and determination of relevant observables is to use
the estimator (\ref{wigEstimator}) to obtain the Wigner functions associated with the two states.
The results are shown in Fig.\ref{Fig5}; as in the previous numerical experiments, in order
to minimize artifacts related to the numerical integration in (\ref{kernel}), a low pass Gaussian
filter has been applied to the raw images.

With these results, one can then estimate the mean values of relevant observables,
as number, position and momentum operators together with the squeezing parameter, obtained from the computation
of the variances of position and momentum. The results are summarized in Table \ref{tabfiltro7}.
The obtained expectation values for the number operator explicitly exhibit the three order of magnitude difference between the two states. Furthermore, these results show that the analyzed experimental states are indeed 
minimum uncertainty states, with vanishing squeezing parameter.
\begin{table}[hbt]
 \caption{\small Estimate of different expectation values for the experimental coherent states
 with optical density $OD=4.5$ and $OD=7$, respectively.}
\begin{center}
\begin{tabular}{| c || c || c |}
  $\hat{O}$ & $E^{h}_{\hat{O}} $ \, ($OD=4.5$) & $E^h_{\hat{O}}$ \, ($OD=7$) \phantom{\bigg|}  \\ \hline
  $\hat{n}$ & $51 \pm 7$ & $0.2 \pm 0.3$ \\	
  $\hat{q}$  & $10 \pm 1$  & $0.7 \pm 0.2$ \\
  $\hat{p}$   & $0.1 \pm 0.2$ & $0.0 \pm 0.2$  \\
  $\sigma_{\hat{q}}^2$  & $0.7 \pm 0.3$ &$0.5 \pm 0.2$ \\
  $\sigma_{\hat{p}}^2$ & $0.6 \pm 0.3$ & $0.5 \pm 0.3$ \\
  $\xi$ & $0.0 \pm 0.2$ & $0.0 \pm 0.2$ \\
\end{tabular}
\end{center}
\label{tabfiltro7}
\end{table}

\section{Discussion}

The previous results demonstrate the effectiveness of the minimax statistical techniques in the
reconstruction of quantum states of light in presence of large electronic noise.
These methods allow to circumvent convergence problems that arise when using standard pattern function methods in estimating observable averages with low detection efficiency, namely $\eta<50\%$ \cite{DAriano1,Kiss1,Herzog,DAriano2,Kiss2,Richter,DAriano3}. Nevertheless, it is interesting to investigate the possible relations
between these two approaches, by replacing the cutoff truncation in the integration with respect to $t$ in (\ref{kernel})
with a suitable Gaussian regularization. In particular
we consider the following alternate kernel functions instead of 
$R^{h,\eta}_{\hat{O}}(Y_i,\Phi_i)$,

\begin{equation}
R^{\varepsilon}_{\hat{O}}(Y_i, \Phi_i)=\int_{\mathbb{R}^2} dq \, dp \, \hat{O}(q,p)\, e^{-\varepsilon(q^2+p^2)}
\, \int_{-\infty}^{+\infty} dt \frac{|t|}{4 \pi} e^{- \imath t (q \cos{\Phi_i} + p \sin{\Phi_i} - \frac{Y_i}{\sqrt{\eta}})}\
e^{t^2 \frac{1-\eta}{4 \eta}}\ ,
\label{kernel2}
\end{equation}
where $\varepsilon$ is a positive regularization parameter.
In principle, this allows for a more direct estimate of the expectation values of any observable,
provided $0< \varepsilon< \eta/(1-\eta)$.
For instance, in the case of position, momentum and number operators, one finds:
\begin{eqnarray}
\label{opq}
R^{\varepsilon}_{\hat{q}}(Y_i, \Phi_i)&=& \int_{-\infty}^{+\infty} dt\,\frac{\cos\Phi_i}{8\varepsilon^2}\,
 t\,|t|\sin\Big(\frac{t\,Y_i}{\sqrt{\eta}}\Big)\, e^{-t^2\, \frac{\eta-\varepsilon(1-\eta)}{4\eta\varepsilon}}\ ,\\
\label{opp}
R^{\varepsilon}_{\hat{p}}(Y_i, \Phi_i)&=& \int_{-\infty}^{+\infty} dt\,\frac{\sin\Phi_i}{8\varepsilon^2}\,
 t\,|t|\,\sin\Big(\frac{t\,Y_i}{\sqrt{\eta}}\Big)\, e^{-t^2\, \frac{\eta-\varepsilon(1-\eta)}{4\eta\varepsilon}}\ ,\\
\label{opn}
R^{\varepsilon}_{\hat{n}}(Y_i, \Phi_i)&=& \int_{-\infty}^{+\infty} dt\,\frac{1}{32\varepsilon^3}\,
 |t|\,\cos\Big(\frac{t\,Y_i}{\sqrt{\eta}}\Big)\,e^{-t^2\, \frac{\eta-\varepsilon(1-\eta)}{4\eta\varepsilon}} [4\varepsilon(1-\varepsilon)-t^2]\ .
\end{eqnarray}
Notice that these quantities converge to functions 
$R^0_{\hat{O}}(Y_i, \Phi_i)$ in the limit $\varepsilon\to 0$:
\begin{eqnarray}
\label{opq0}
&&\hskip -1cm
R^0_{\hat{q}}(Y_i,\Phi_i)=\frac{2}{\sqrt{\eta}}Y_i\cos\Phi_i\ ,\\
\label{opp0}
&&\hskip-1cm
R^0_{\hat{p}}(Y_i,\Phi_i) =\frac{2}{\sqrt{\eta}}Y_i\sin\Phi_i\ ,\\
\label{opn0}
&&\hskip-1cm
R^0_{\hat{n}}(Y_i,\Phi_i)= \Big(\frac{Y_i}{\sqrt{\eta}} \Big)^2-\frac{1}{2 \eta}\ .
\end{eqnarray}
These expressions coincide with the so-called kernel functions of the standard pattern
function based tomographic techniques \cite{DAriano3}. Unfortunately, the limit $\varepsilon\to 0$ can not be taken in the case of more general observables with higher powers of $\hat q$ and $\hat p$ since the integrals diverge as powers of $1/\varepsilon$. This fact may be related to the mentioned convergence problems of the standard pattern function method. 

It is important to note that the estimation of the mean values of $\hat q$, $\hat p$ and $\hat n$ through the use of
the limit kernel functions (\ref{opq0})-(\ref{opn0}) is not sufficient to completely characterize any quantum states, like coherent and squeezed states.

\section{Conclusions}
Here we report the characterization of a time domain balanced homodyne detection apparatus operating in presence of large electronic noise corresponding to an overall detection efficiency $\eta=0.31$. We used the detector combined with tomographic reconstruction techniques for discriminating between different quantum states of light. A minimax adaptive reconstruction of the Wigner function
has been adopted. This approach allows us to circumvent possible convergence problems arising from low detector efficiency in the standard pattern function based
quantum tomography. 

The effectiveness of such a method has been verified in two ways.
At first, we  calculated the Wigner function  for simulated data of coherent and squeezed states and, then, for real experimental
homodyne data of coherent states with different mean photon numbers. In all cases it is proved that it is possible to efficiently
reconstruct the associated Wigner function, asserting the Gaussian character of the quantum states, and evaluating their
relevant parameters.

The present study demonstrates that even low efficient ($\sim30\%$) homodyne detectors can be usefully employed to
study the nature of quantum states of light provided that non-standard statistical tools as the minimax 
methods are used to reconstruct their Wigner functions. These results may be important whenever quantum optics
techniques are employed to investigate the dynamics of out of equilibrium states and the presence of quantum coherence in condensed matter.

\section*{Acknowledgments}
Useful discussions with Simone Cialdi and Marco Barbieri are acknowledged. This work has been supported by MIUR (FIRB "LiCHIS"-RBFR10YQ3H), University of Trieste ("FRA 2009" and "FRA 2011") and  European Union, Seventh Framework Programme, under the project GO FAST, grant agreement no. 280555.

\appendix
\section{Homodyne detection in the pulsed regime: multimode treatment}

Homodyne detection in a pulsed regime requires a formal generalization of its theoretical description with respect to the one-mode regime; indeed, the \emph{LO} and the \emph{signal} at the beam splitter are not monochromatic.
Classically, the electric field of a pulsed laser beam can be represented as a mode-locked superposition of amplitudes:
\begin{eqnarray}\label{d1}
  E(t) = \sum_{l=-M}^{M} |\alpha_l| \, e^{i\,\Phi_l (t)}\ ,\qquad \Phi_l (t) = \omega_l \, t + \varphi_l\ ,
\end{eqnarray}
where the phases $\Phi_l (t)$ are mode-locked by the condition $\varphi_l = l \, \varphi_0$, with $\varphi_0$ a reference phase. Each frequency $\omega_l$ contributes to the field with the amplitude $\alpha_l = |\alpha_l| \,  e^{i \Phi_l}$ and the number of contributing frequencies depends on the shape of the pulse.
Quantized pulsed laser light is described by associating to each monochromatic component a coherent state $\ket{\alpha_l}$, that is an eigenstate of the annihilation operator $\hat{a}_l$  of photons in the mode of frequency $\omega_l$, $\hat{a}_l\ket{\alpha_l}=\alpha_l\ket{\alpha_l}$,
and to the entire pulse the tensor product
\begin{eqnarray}\label{d5}
\ket{\bar{\alpha}} = \bigotimes_{l=-M}^{M} \ket{\alpha_l},
\end{eqnarray}
where $\bar{\alpha}$ is the vector whose components are the amplitudes $\alpha_l$.
By means of the creation and annihilation operators $\hat{a}_l$ and $\hat{a}^{\dagger}_l$  each monochromatic coherent state reads
\begin{eqnarray}\label{d2}
\ket{\alpha_l} = D(\alpha_l) \ket{0}\ ,\qquad D(\alpha_l)= e^{\alpha_l \hat{a}^{\dagger}_l- \alpha^{\ast}_l \hat{a}_l}\ ,
\end{eqnarray}
where $\ket{0}$ is the vacuum state and $D(\alpha)$ is the so-called displacement operator.\\
Since the creation and annihilation operators pertaining to different modes commute, the pulsed coherent state (\ref{d5}) can be conveniently recast as:
\begin{eqnarray}
\label{d51}
 \ket{\bar{\alpha}} =D(\bar\alpha)\ket{0}\ ,\qquad D(\bar\alpha)=e^{\hat{A}^\dag(\bar{\alpha})-\hat{A}(\bar{\alpha})}\ ,
\end{eqnarray}
by mean of a displacement operator $D(\bar\alpha)$ expressed in terms of multi-mode operators
\begin{eqnarray}
\label{ca}
\hat{A}^{\dagger} (\bar{\alpha}) = \sum_l \alpha_l \, \hat{a}^{\dagger}_l\ ,\qquad \hat{A}(\bar{\alpha}) = \sum_l \alpha^{\ast}_l \hat{a}_l\ .
\end{eqnarray}
The reason for labeling the pulsed coherent state by $\ket{\bar{\alpha}}$ can now be easily understood.
The state of one photon of frequency $\omega_l$ is given by  $\hat{a}^{\dagger}_l \ket{0} = \ket{1_l}$, while a generic non-monochromatic superposition of frequencies $\omega_l$ with amplitudes $\alpha_l$ corresponds to the state $\ket{1_{\bar{\alpha}}} = \sum_l \alpha_l \ket{1_l}$ that results from applying the multimode operator
$\hat{A}^{\dagger}(\bar{\alpha})$ to the vacuum state:
\begin{eqnarray}
\label{d6}
  \hat{A}^{\dagger} (\bar{\alpha}) \ket{0} = \sum_l \alpha_l \, \hat{a}^{\dagger}_l \ket{0} = \sum_l \alpha_l \, \ket{1_l} = \ket{1_{\bar{\alpha}}}\ .
\end{eqnarray}
Therefore, $\hat{A}^{\dagger}(\bar{\alpha})$ is the creator operator of a single-photon in the (not normalized) superposition state $\ket{1_{\bar\alpha}}$, while
$\hat{A}(\bar{\alpha})$ destroys a photon in the same state;
thus, the quantum state of the pulsed laser is a coherent state associated not with a single amplitude $\alpha_l$, but with the
vector $\bar{\alpha}$ of all the amplitudes contributing to the pulse: in other words, we have a Poissonian distribution not with respect to the number of photons in a monochromatic wave, but to the number of photons in the superposition $\ket{1_{\bar{\alpha}}}$.\\
The normalized operators
\begin{eqnarray}
\label{new}
\hat{A}=\frac{\hat{A}(\bar\alpha)}{|\bar\alpha|}\ ,\qquad \hat{A}^\dag=\frac{\hat{A}(\bar\alpha)}{|\bar\alpha|}\ ,
\end{eqnarray}
where $|\bar\alpha|^2=\braket{1_{\bar\alpha}\vert1_{\bar\alpha}}$, satisfy the canonical commutation relations
\begin{eqnarray}\label{d7}
\left[ \hat{A}, \hat{A}^{\dagger}\right]=\frac{1}{|\bar\alpha|^2} \sum_{ij} {\alpha}^{\ast}_i \alpha_j \,  [\hat{a}_i, \hat{a}^{\dagger}_j] = 1 \  .
\end{eqnarray}
By means of them  we can now extend the state-tomography techniques to the case in which the \emph{signal} and the \emph{LO} are pulsed.
In the monochromatic case, when the \emph{signal} mode $a$ interferes with the \emph{LO} mode $b$ on the beam splitter, the photo-current operator is
$\hat{I} =  \hat{a}^{\dagger} \hat{b} + \hat{b}^{\dagger} \hat{a}$ \cite{Ferraro}.\\
If more frequencies $\omega_l$ are present both in the \emph{LO} and in the \emph{signal}, each one of the corresponding mode
operators will be subjected to the beam splitting transformation and the detectors will ideally register photons of all involved frequencies. Then, the photo-current operator becomes
$\hat{I} =  \sum_l  \hat{a}^{\dagger}_l \hat{b}_l + \hat{b}^{\dagger}_l\hat{a}_l$.

If the \emph{LO} is in a pulsed coherent state $\ket{\bar{z}} = e^{\hat{B}^{\dagger}(\bar{z}) - \hat{B} (\bar{z})} \ket{0}$, with generalized creation and annihilation operators $\hat{B}^{\dagger}(\bar{z}) = \sum_l z_l \hat{b}^{\dagger}_l$ and $\hat{B}(\bar{z}) = \sum_l z^{\ast}_l \hat{b}_l$,
the phase difference $\Phi$ between the \emph{LO} and the \emph{signal} is changed by the action of the piezoelectric translator placed in the \emph{LO} arm on all the \emph{LO} modes:
\begin{eqnarray}\label{BSBSA}
\hat{b}_l \rightarrow  \hat{b}_l \, e^{i \Phi}\ ,\qquad
\hat{b}^{\dagger}_l  \rightarrow \hat{b}^{\dagger}_l \, e^{-i \Phi}\ .
\end{eqnarray}
The photo-current operator which is measured by the pulsed homodyne setup is thus given by
\begin{eqnarray}
\label{photocurrent}
\hat{I}_\Phi=\sum_l\left(\hat{a}_l^\dag\,\hat{b}_l\,\, e^{i \Phi}\,+\, \hat{a}_l\,\hat{b}_l^\dag \, e^{-i \Phi}\right)\ .
\end{eqnarray}
Let $\hat{\rho}_s$ denote the quantum  state (density matrix) of the \emph{signal} field and by $\ket{\bar z}\bra{\bar z}$ the projector onto the (coherent) state of the incoming pulse. Then, using that $\hat{b}_l\ket{\bar{z}} = z_l\ket{\bar z}$ and the expressions in  (\ref{ca}), the  expectation value $I_\Phi= {\rm Tr}\left(\rho_s\otimes\ket{\bar z}\bra{\bar z}\,\hat{I}_\Phi\right)$ of the photo-current is calculated as follows:
\begin{eqnarray}
\nonumber
I_{\Phi}&=&
\sum_{l} \left( {\rm Tr} \left[\hat{\rho}_s  \, \hat{a}^{\dagger}_l \right] \cdot \braket{\bar{z}|\hat{b}_l|\bar{z}} \, e^{i \Phi} + h.c. \right)\\
\label{tracciaPuls1}
& = & \sum_{l} \left( {\rm Tr} \left[\hat{\rho}_s  \, (\hat{a}^{\dagger}_l \, z_l \, e^{i \Phi} +  \hat{a}_l \, {z}^{\ast}_l \, e^{-i \Phi})\right]\right) \nonumber \\
           & = &  {\rm Tr} \left[\hat{\rho}_s \, \left(\hat{A}^{\dagger} (\bar{z}) \, e^{i \Phi} + \hat{A} (\bar{z}) \, e^{-i \Phi}\right)\right]\ .
\end{eqnarray}
Using  (\ref{new}) and comparing these results in the standard treatment of homodyne detection for the single-mode case, one realizes that in the pulsed regime one measures a quantity $I_\Phi$,
\begin{equation}
\label{mean}
I_\Phi=\sqrt{2}\,|\bar z|\,{\rm Tr}\left[\hat{\rho}_s\,\hat{X}_\Phi\right]\ ,
\end{equation}
proportional to a quadrature which generalizes that in (\ref{quad}):
\begin{equation}
\label{Quad}
\hat{X}_\Phi=\frac{\hat{A}^{\dagger}\, e^{i \Phi} + \hat{A}\, e^{-i \Phi}}{\sqrt{2}}\ .
\end{equation}
For this reason, in the main text of the letter we have limited the discussion to the simpler case of a single mode homodyne detection, with the proviso that whenever a quadrature is used, it actually refers to its expression (\ref{Quad})
in the pulsed regime.
In particular, the measured data can be used to reconstruct the expectation values of all \emph{signal} observables that can be expressed as functions of the operators $\hat{A}$ and $\hat{A}^\dag$. For instance, the mean value of the photo-current operator second moment with respect to the \emph{LO} pulsed coherent state $\ket{\bar z}\bra{\bar z}$ is
\begin{eqnarray}
\bra{\bar z}\,\hat{I}_\Phi^2\ket{\bar z}
=2\,|\bar z|^2\,\hat{X}^2_\Phi\,+\,\sum_l\hat{a}^\dag_l\hat{a}_l\ ,
\label{variance0}
\end{eqnarray}
and differs from the pulsed quadrature second moment $\hat{X}^2_\Phi$ by the number of photons in the pulsed \emph{signal} which is to be taken much smaller than
the intensity $|\bar z|^2$ of the pulsed \emph{LO}. A similar suppression by $|\bar z|^{-2}$ occurs for the correction terms
appearing in higher moments so that the distribution of the outcomes of the homodyne photocurrent is equal to that of the corresponding field quadratures. As a final remark, notice that, in the case of only one frequency mode, the above setting reduces to the standard homodyne tomographic one.

\end{document}